\documentclass{article}

\usepackage{arxiv}
\usepackage{graphicx}
\usepackage{lscape}

\usepackage[utf8]{inputenc} 
\usepackage[T1]{fontenc}    
\usepackage{hyperref}       
\usepackage{url}            
\usepackage{booktabs}       
\usepackage{amsfonts}       
\usepackage{nicefrac}       
\usepackage{microtype}      
\usepackage{lipsum}

\title{Testing for time-varying properties under misspecified conditional mean and variance}

\author{
  Daiki Maki\thanks{This research was supported by KAKENHI (Grant number: 16K03604). Address: Faculty of Commerce, Doshisha University, 
         Karasuma-higashi-iru, Imadegawa-dori, Kamigyo-ku, Kyoto Japan 602-8580 
         (E-mail:dmaki@mail.doshisha.ac.jp)}  \\ 
         Doshisha University
         \and \textbf{Yasushi Ota} \thanks{This research was supported by KAKENHI (Grant number: 18K03439). Faculty of Management, Okayama University of Science, 1-1 Ridaicyou, Okayama City, Okayama Japan 700-0005 
(E-mail:yota@mgt.ous.ac.jp)} \\ Okayama University of Science}

\begin{document}
\maketitle

\begin{abstract}
This study examines statistical performance of tests for time-varying properties under misspecified conditional mean and variance. 
When we test for time-varying properties of the conditional mean in the case in which data have no time-varying mean but have time-varying variance, 
asymptotic tests have size distortions. 
This is improved by the use of a bootstrap method. 
Similarly, when we test for time-varying properties of the conditional variance in the case in which data have time-varying mean but no time-varying variance, 
asymptotic tests have large size distortions. 
This is not improved even by the use of bootstrap methods.
We show that tests for time-varying properties of the conditional mean by the bootstrap are robust regardless of the time-varying variance model, 
whereas tests for time-varying properties of the conditional variance do not perform well in the presence of misspesified time-varying mean.
\end{abstract}


\section{Introduction}
Many economic and financial time series data have time-varying properties. 
Their properties are roughly classified into two types. 
One is a property about the conditional mean. 
Constant and/or autoregressive parameters change with time as time-varying properties for the mean. 
Another time-varying property is the conditional variance of the error term for a regression model. 
Variance of the error term, that is, volatility, is frequently modeled by the autoregressive conditional heteroskedasticity (ARCH). 
As introduced by Dahlhaus and Rao (2006), a property of the time-varying model for volatility is that the parameters of volatility models change with time. 
Amado and Ter\"{a}svirta (2013) and Kim and Kim (2016) recently provided a time-varying volatility model and its applications. 

Linearity tests are usually used to investigate time-varying properties. 
When we test for time-varying properties of the conditional mean, 
we assume a homoskedastic variance or a correctly specified linear variance process for the error term of the condtional mean. 
When we test for time-varying properties of the conditional variance, 
we assume a correctly specified linear process for a mean. 
However, it is difficult to know which time-varying property for the mean or variance is present a priori. 
This challenge implies that researchers may erroneously test for time-varying properties of the mean, 
although volatility actually has time-varying properties. 
Similarly, researchers might erroneously test for time-varying properties of the variance, 
although the mean actually has time-varying properties. 
In fact, Pitarakis (2004) and Perron and Yamamoto (2019) showed that a test for a change in regression coefficients (variance) 
ignoring or misspecifying the presence of change in variance (regression coefficients) 
causes poor statistical properties. 
For the influence of ARCH on inference of misspecified models, Lumsdaine and Ng (1999) showed over-rejection of ARCH tests in the presence of misspecified conditional mean models. 
Van Dijk, Franses, and Lucas (1999) demonstrated size distortions of ARCH tests when a process has additive outliers. 
Balke and Kapetanios (2007) pointed out that supurious ARCH effects appear when nonlinearity of the mean is ignored. 
If we were to perform erronous tests and fail to obtain reliability from the derived results, 
the correct model construction and evaluation are difficult. 
Accordingly, it is important to clarify the influence of tests on the misspecified models.

This study examines the influence of time-varying tests on misspecified conditional mean and variance.  
In particular, we clarify the statistical performance of tests for time-varying variance when a process has time-varying mean with homoskedastic variance. 
We also investigate statistical performance of tests for time-varying mean when a process has time-varying variance with a linear mean model. 
Some studies, including Lumsdaine and Ng (1999), van Dijk, et al. (1999), and Balke and Kapetanios (2007), investigated problems of misspecified models in ARCH tests. 
However, previous studies have not clarified the influence of time-varying tests on misspecified conditional mean and variance. 
As mentioned above, it is important to clarify time-varying properties correctly. 
This study uses a logistic smooth transition function to model time-varying mean and variance. 
This model smoothes threshold or structural break models in time.  
The time-varying tests used in this study are based on the method introduced by Luukkonen, Saikkonen, and Ter\"{a}svirta (1988a). 
They depeloped linearity tests using a Taylor series approximation to overcome the identification problem pointed out by Davies (1977, 1987).

The simulation results in this study provide evidence that the asymptotic test for time-varying mean has size distortions 
when the conditional variance model is misspecified. 
However, the wild bootstrap method introduced by Liu (1988) improves the size distortions. 
In fact, Becker and Hurn (2009) provided evidence that the wild bootstrap improves size properties when the conditional mean is being tested. 
We can test for time-varying mean by using the wild bootstrap without depending on the form of volatility. 
When we test for time-varying variance in the presence of a misspecified conditional mean, 
asymptotic tests have large size distortions. 
The properties are not improved by the use of the bootstrap ARCH test proposed by Gel and Chen (2012) and the wild bootstrap. 
The results show that the wild bootstrap tests for time-varying mean are robust regardless of the misspecified conditional variance, 
whereas tests for time-varying variance do not perform well in the presence of misspecified conditional mean.

The remainder of this paper is organized as follows. 
Section 2 presents the tests for time-varying mean and variance.  
Section 3 provides statistical properties of the tests for time-varying mean and variance in the presence of misspecified conditional models. 
Finally, Section 4 summarizes and concludes.

\section{Tests for time-varying mean and variance}
We assume the following regression model to test for time-varying mean$^1$. 
\begin{equation}
             y_t=\alpha_0 + \beta_0 y_{t-1} + (\alpha_1 + \beta_1 y_{t-1}) F (t, \ \gamma, \ c)+ u_{t}, 
\end{equation}
where $u_t \sim N(0, \sigma^2)$ and $F(t, \ \gamma, \ c)$ is a transition function to model time-varying properties of (1). 
(1) has a time-varying constant and slope for $y_{t-1}$ depending of the value of $F(t, \ \gamma, \ c)$. 
Lin and Ter\"{a}svirta, T. (1994) provided the asymptotic theory for time-varying regression models.  
(1) reduces to a linear autoregressive model for $F(t, \ \gamma, \ c)=0$ and/or $\alpha_1=0$ and $\beta_1= 0$. 
$F(t, \ \gamma, \ c)$ is denoted as 
\begin{equation}
             F(t, \ \gamma, \ c)= (1+ \exp \{ - \gamma (t-c) \})^{-1}-\frac{1}{2},
\end{equation}
where $\gamma$ is a parameter of smoothness and $c$ is a threshold parameter.   
We assume $\gamma>0$ and $c>0$. 
$F(t, \ \gamma, \ c)$ is bounded between $-1/2$ and $1/2$. 
$F(t, \ \gamma, \ c)$ moves toward $-1/2$ when $t < c$ and small $\gamma (t-c)$,
and moves toward $1/2$ when $t > c$ and large $\gamma (t-c)$. 
$F(t, \ \gamma, \ c)$ takes 0 for $t=c$. 
$F(t, \ \gamma, \ c)$ with $\gamma=\infty$ equals a structural break model, 
because $F(t, \ \gamma, \ c)$ with $\gamma=\infty$ becomes the indicator function that takes the value -1/2 or 1/2. 
For example, Figures 1 and 2 demonstrate the values of $F(t, \ \gamma, \ c)$. 
While Figure 1 depicts the graph of $F(t, \ \gamma, \ c)$ with $T=200$, $\gamma=(0.01,0.1)$, and $c=T/2$, 
Figure 2 depicts it with $T=1,000$, $\gamma=(0.01,0.1)$, and $c=T/2$.   
Figure 1 shows that $F(t, \ \gamma, \ c)$ with $\gamma=0.01$ is almost linear, whereas $F(t, \ \gamma, \ c)$ with $\gamma=0.1$ has a smooth change around $T=100.$ 
For Figure 2, $F(t, \ \gamma, \ c)$ with $\gamma=0.01$ has a smooth change around $T=500$. 
By contrast, $F(t, \ \gamma, \ c)$ with $\gamma=0.1$ has a rapid chagnge around $T=500.$ 
The time-varying properties of (1) depend on the values of $t$, $\gamma$, and $c$.

The null and alternative hypotheses to test for time-varying mean in (1) are denoted as follows
\begin{equation}
H_0: \gamma=0, \ \ \ \ \ H_1: \gamma>0.
\end{equation}
(1) becomes a simple autoregressive (AR) model under $H_0$ and has a time-varying constant and slope under $H_1$ with $\alpha_1$ and $\beta_1$. 
Note that we cannot test for $\gamma=0$ directly, 
because the null hypothesis has an identification problem about $\alpha_1$ and $\beta_1$, 
which are identified only under the alternative hypothesis with $\gamma >0$. 
Davies (1977, 1987) discussed the identification problem whereby a nuisance parameter is present only under the alternative hypothesis. 
A solution for the identification problem is to use the Taylor series approximation introduced by Luukkonen, Saikkonen, and Ter\"{a}svirta (1988a).

When we use a Taylor series approximation around $\gamma=0$, 
(1) with $F(t, \gamma, c)$ is given by 
\begin{equation}
       y_t=\phi_0 + \phi_1 y_{t-1} + \phi_2 t + \phi_3 t y_{t-1} + e_{t}, 
\end{equation}
where $e_t$ is an error term for the new regression model. 
We can formulate the null and alternative hypotheses to test for time-varying mean in (4) as follows.
\begin{equation}
H_0: \phi_2=\phi_3=0, \hspace{5ex}  H_1: H_0 \hspace{1ex} \mbox{is not true}.
\end{equation} 
We rewrite (4) as 
\begin{equation}
y_t=\Phi^{\prime} Y_t +e_t, 
\end{equation}
where $\Phi=(\phi_0, \phi_1, \phi_2, \phi_3)^{\prime}$ and $Y_t=(1, y_{t-1}, t, t y_{t-1})^{\prime}$. 
The usual test for (5) uses the following Wald statisic. 
\begin{equation}
M_a=\frac{1}{\hat{\sigma}^2}\hat{\Phi}_2^{\prime} \Bigg [R \bigg (\sum_{t=1}^T Y_t Y_t^{\prime} \bigg)^{-1} R^{\prime} \Bigg]^{-1} \hat{\Phi_2}, 
\end{equation}
where $\hat{\Phi_2}$ is the estimate of $\Phi_2=(\phi_2, \phi_3)^{\prime}$, $\hat{\sigma}^2$ is the estimate of the residual variance obrained from (4), 
and $R$ is the matrix such that $R\hat{\Phi}=\hat{\Phi}_2$.   
$M_a$ has an $F$ distribution with $(2, T-4)$ degrees of freedom under the null hypothesis.

Standard asymptotic tests for linearity often cause spurious nonlinearity when errors have heteroskedastic variance, including ARCH, generalized ARCH (GARCH), stochastic volatility, and structural breaks. 
A method to relieve the influence is to use heteroskedasticity consistent covariance matrix estimator (HCCME) introduced by White (1980). 
This is a popular method to test for linearity in the presence of heteroskedasticity, 
since test statistics based on the HCCME asymptotically have the same distribution as the original test statistics under the null hypothesis.  
However, linearity tests using HCCME do not perform well under the null hypothesis of linearity with ARCH and GARCH type errors particularly for finite samples. 
HCCME cannot improve the influence of heteroskedastic variance on inference of the conditional mean sufficiently. 
As pointed out by Pavlidis, Paya, and Peel (2013), the presence of ARCH errors complicate tests for lineality because ARCH errors are like nonlinear processes for conditional mean. 
Some studies, including Pavlidis, Paya, and Peel (2010) and Maki (2014), demonstrated this problem. 
An better method is the wild bootstrap proposed by Liu (1988). 
This can resample data with unknown heteroskedastic variance and yeild better statistical performance than the asymptotic and the HCCME tests can.   
The wild bootstrap test for time-varying mean takes the following procedure. 
\\
Step 1. Estimate regression model (4) and compute test statistic (6). 
\\
Step 2. Estimate the regression model under the null hypothesis with $\phi_2=\phi_3=0$ and obtain the parameter estimates $\hat{\phi}_0$ and $\hat{\phi}_1$ and residuals denoted as $\hat{e}_{0t}$. 
\\
Step 3. Generate the bootstrapped sample as follows: 
\begin{equation}
y_t^{\ast}=\hat{\phi}_0+\hat{\phi}_1 y_{t-1} + e_{t}^{\ast}, 
\end{equation}
where $e_{t}^{\ast}=\epsilon_{t} \hat{e}_{0t}$. 
We set $\epsilon_{t}$ to independent and identically distributed $N(0,1)$$^2$. 
\\
Step 4. Estimate regression model (4) using the generated bootstrap sample (7) and compute test statistic (6) denoted as $M_{wb}$. 
\\
Step 5. Repeat steps 3 and 4 many times. 
\\
Step 6. Compute the bootstrap $p$-value as follows:
\begin{equation}
P(M_{wb})=\frac{1}{M}\sum_{m=1}^M I(M_{wb} > M_a), 
\end{equation}
where $M$ is the number of bootstrap iterations and 
$I(\cdot)$ is an indicator function such that $I(\cdot)$ is 1 if $(\cdot)$ is true and 0 otherwise. 
Usually, more than 1,000 times are enough for the number of bootstrap iterations. 
Andrews and Buchinsky (2000) and Davidson and MacKinnon (2000) discussed the problem of the number of bootstrap iterations. 
The null hypothesis is rejected if (8) is smaller than a significant level.

We next explain the test for time-varying smooth transition variance. 
Consider the following data generating process (DGP). 
\begin{eqnarray}
             y_t=\alpha_0 + \beta_0 y_{t-1} + u_{t}, \\ 
             u_t=h_t \epsilon_t, \\
             h_t^2=a_0+b_0 u_{t-1}^2+ (a_1+b_1 u_{t-1}^2)F(t, \ \gamma, \ c), \\
             F(t, \ \gamma, \ c)= (1+ \exp \{ - \gamma (t-c) \})^{-1}-\frac{1}{2}.   
\end{eqnarray}
(9) has a linear AR model for mean and time-varying smooth transition ARCH process (11) for variance. 
While Hagerud (1996) proposed the smooth transition GARCH model with the transition variable $u_{t-1}$, 
(12) has the transition variable $t$. 
The transition function for time-varying ARCH is similar to that of time-varying mean$^{3}$. 
Since (11) also has an identification problem, we use a Taylor series approximation around $\gamma=0$ in order to test for time-varying smooth transition ARCH. 
(11) is rewritten as 
\begin{equation}
 h_t^2=\rho_0+\rho_1 u_{t-1}^2+ \rho_2 t + \rho_3 t u_{t-1}^2 +\upsilon_t, 
\end{equation}
where $\upsilon_t$ is an error term, including white noise and the remainder term from the Taylor approximation. 
The null hypothesis and alternative hypothesis of test for time-varying ARCH are given by
\begin{equation}
H_0: \rho_1=\rho_2=\rho_3=0, \ \ \ \ \ H_1: H_0 \ \mbox{is not true}.
\end{equation} 
(14) nests ARCH test$^{4}$. 
In order to test for (14), we first estimate (9) and obtain residuals. 
The test statistic is  
\begin{equation}
V_a=\frac{(SSR(h)_0-SSR(h)_1)/3}{SSR(h)_1/(T-4)},  
\end{equation}
where $SSR(h)_0$ is the sum of the squared residuals obtained from the estimation of (13) with $\rho_1=\rho_2=\rho_3=0$ and  
$SSR(h)_1$ is the sum of the squared residuals obtained from the estimation of (13).  
$V_a$ has an $F$ distribution with $(3, T-4)$ degrees of freedom under the null hypothesis$^{5}$.

Gel and Chen (2012) introduced a new bootstrap test for ARCH to improve the size and power of the asymptotic test. 
We use their approach to test for time-varying ARCH. 
The procedure is denoted as follows. 
\\
Step 1. Estimate regression model (13) and compute test statistic (15). 
\\
Step 2. Estimate the regression model under the null hypothesis with $\rho_1=\rho_2=\rho_3=0$ and obtain the parameter estimates $\hat{\rho}_0$ and residuals denoted as $\hat{\upsilon}_{0t}$. 
\\
Step 3. Generate the bootstrapped sample as follows: 
\begin{equation}
h_t^{\ast}=\hat{\rho}_0+ \upsilon_{t}^{\ast}, 
\end{equation}
where $\upsilon_{t}^{\ast}$ is randomly selected from $(\hat{\upsilon}_{01} \cdots \hat{\upsilon}_{0T})$. 
\\
Step 4. Estimate regression model (13) using the generated bootstrap sample (16) 
and compute test statistic (15) denoted as $V_{b}$. 
\\
Step 5. Repeat steps 3 and 4 many times. 
\\
Step 6. Compute the bootstrap $p$-value as follows:
\begin{equation}
P(V_{b})=\frac{1}{M}\sum_{m=1}^M I(V_{b} > V_a). 
\end{equation}
The null hypothesis is rejected if the $p$-value (17) is smaller than a significant level. 
We also use the wild bootstrap test for (14). This is almost smilar to the bootstrap test shown above. 
The main difference is Step 3. 
The wild bootstrap approach replace $\upsilon_{t}^{\ast}$ as $\tilde{\upsilon}_{t}=\hat{\upsilon}_{0t} \eta_t$, where $\eta_t \sim N(0,1)$. 
We denote the wild bootstrap test for time-varying ARCH as $V_{wb}$. 
The $p$-value is given by 
\begin{equation}
P(V_{wb})=\frac{1}{M}\sum_{m=1}^M I(V_{wb} > V_a). 
\end{equation}

\section{Statistical properties of time-varying tests}
This section examines the statistical properties of the time-varying tests reviewed in section 2. 
We conduct Monte Carlo simulations to compare the size and power of the test statistics. 
The simulations are based on 10,000 replications, the nominal level at 5\%, and sample sizes with $T=100$, $200$, $400$, and $1000$. 
We show only nominal power properties and do not present size-corrected power 
because the purpose of the paper is to compare the tests and usual applied practitioners do not use size-corrected tests.
Bootstrap tests have 1,000 replications.  
In order to avoid the effect of initial conditions, data with $T+100$ are generated. 
The initial 100 samples are discarded and we use the data with sample size $T$. 
Tests compared in this section are denoted as $M_a$, $M_{wb}$, $V_a$, $V_{b}$, and $V_{wb}$.

First, as a benchmark, autoregressive processes with homoskedastic error are generated. 
\begin{equation}
             y_t=\alpha_0 + \beta_0 y_{t-1} + u_{t}, \\
\end{equation}
where  $u_{t} \sim \mbox{i.i.d.}N(0,1)$. 
$\alpha_0$ is set to $\alpha_0=1$. 
The persistent parameter $\beta_0$ is set to $\beta_0=0.3$ and $0.9$ in order to investigate the influence of persistence of the process on time-varying tests. 
Since (19) has no time-varying properties in mean and variance, 
the rejection frequencies of each test demonstrate the empirical size for (19). 
Table 1 presents the rejection frequencies of each test.  
When the persistent parameter is $\beta_0=0.3$, $V_a$ has slight under-rejection for $T=100$, $200$, and $400$. 
Other tests have the empirical size near to 5\% nominal level and reasonable size properties regardless of sample size. 
Although time-varying tests for mean, $M_a$ and $M_{wb}$, have over-rejection in small samples with $\beta_0=0.9$, 
the over-rejection decreases with the increased sample size. 
Unlike time-varying tests for mean, time-varying tests for variance do not have size distortions, except for $V_a$ with $T=100$, 
which has slight under-rejection.

Next, we consider the following DGP to examine the power properties of $M_a$ and $M_{wb}$ 
and the size properties of $V_a$, $V_b$, and $V_{wb}$ under the following time-varying mean model. 
\begin{eqnarray}
             y_t=1 + 0.3 y_{t-1} + (\alpha_1 + \beta_1 y_{t-1}) F (\cdot)+ u_{t}, \\
             F(\cdot)= (1+ \exp \{ - \gamma (t-c) \})^{-1}-\frac{1}{2}. 
\end{eqnarray}
Parameters for the time-varying mean are set to $(\alpha_1,\beta_1)=(0,0.3),$ $(0,0.6),$ $(0.5,0.3),$ and $(1,0.3)$, respectively.  
$F(\cdot)$ moves from $-1/2$ to $1/2$. 
For $(\alpha_1,\beta_1)=(1,0.3)$, (20) has time-varying properties from $y_t=0.5 + 0.15 y_{t-1} + u_{t}$ to $y_t=1.5 + 0.45 y_{t-1} + u_{t}$. 
We examine the effect of the magnitude of change for constant and AR parameters on performance of the tests. 
Smoothness parameter $\gamma$ in $F(\cdot)$ is set to $\gamma=(0.01,0.1)$. 
$\gamma=0.01$ means more smooth and slight change in $F(\cdot)$ than $\gamma=0.1$. 
In addition, the threshold parameter $c$ is set to $c=T/2^{6}$. 
The simulation results are shown in Table 2.  
The powers of $M_a$ and $M_{wb}$ are low for parameter sets $(\alpha_1,\beta_1,\gamma)=(0,0.3,0.01)$ with $T=100$ and $200$. 
They increase rapidly with the large sample size. 
We can observe that the magnitude of $\alpha_1$ and $\beta_1$ affects the powers of $M_a$ and $M_{wb}$. 
For example, the powers of $M_a$ and $M_{wb}$ for $(\alpha_1,\beta_1)=(0,0.3)$ with $T=200$ are 0.134 and 0.132, respectively, 
whereas those for  $(\alpha_1,\beta_1)=(1,0.3)$ with $T=200$ are 0.694 and 0.703, respectively. 
When the smoothness parameter is $\gamma=0.1$, $M_a$ and $M_{wb}$ have sufficient power.  
$\gamma$ has clear impact on the performance of $M_a$ and $M_{wb}$. 
Generally, we observe that $M_a$ and $M_{wb}$ have ability to find time-varying mean properties in the presence of homoskedastic variance.

For $V_a$, $V_b$, and $V_{wb}$, the results in Table 2 show the empirical size properties of the tests, because (20) has homoskedastic variance. 
$V_a$ performs well in small samples. 
It has over-rejection for $\gamma=0.01$ and $T=1,000$. 
In particular, $V_a$ has a rejection frequency 0.325 and large size distortions when $(\alpha_1, \beta_1)=(0, 0.6)$ and $T=1,000.$ 
We find that the increase in the time-varying AR parameter causes high rejection frequencies for the time-varying variance test under the null hypothesis of homoskecastic variance.  
Although the change of a constant of (20) also increases the rejection frecencies, 
the effect is not larger than the increase in the AR parameter. 
The rejection frecencies of $V_a$ for $T=1,000$ with $(\alpha_1,\beta_1)=(0,0.3)$ and $(\alpha_1,\alpha_1)=(1,0.3)$ are 0.068 and 0.119, respectively.  
Similar properties are observed for $V_b$. 
The results imply that the bootstrap time-varying ARCH test does not perform well under the null hypothesis of homoskedastic variance with a misspecified conditional mean. 
By contrast, the size properties of $V_{wb}$ outperform those of $V_a$ and $V_b$. 
$V_{wb}$ does not have over-rejection even for $T=1,000$. 
However, $V_{wb}$ has size distortions for $\gamma=0.1$.  
Actually, the rejection frequencies of $V_{wb}$ for $T=1,000$ with $(\alpha_1,\beta_1)=(0,0.6)$ and $(\alpha_1,\alpha_1)=(1,0.3)$ are 0.253 and 0.113, respectively. 
$V_{wb}$ does not improve the size properties when the DGP has time-varying mean with homoskedastic error.  
$V_a$ and $V_b$ for $\gamma=0.1$ have larger size distortions than those for $\gamma=0.01$.

When we test for time-varying ARCH in the presence of a misspecified conditional mean, 
the tests cannot lead to reliable results, particularly for a large change in time-varying parameters with a large smoothness parameter $\gamma.$ 
As shown by Balke and Kapetanios (2007), neglected nonlinearity in the conditional mean cause spurious heteroskedasticity 
because the conditional variance includes extra nonlinearity and lags. 
This is more clear when $T$ increases$^{7}$.  
If the mean is misspeficied, 
the residual includes nonlinearity of the mean. 
Large $b_1$ and sample size lead to stronger nonlinearity. 
This is the reason for poor performance of the time-varying variance test.  
Therefore, time-varying ARCH tests in the presence of a misspecified conditional mean have size distortions. 

Table 3 reports rejection frequencies of the tests for the AR process with ARCH error. 
The DGP is the following. 
\begin{eqnarray}
             y_t=1 + 0.3 y_{t-1} + u_{t},  \\
             u_t =h_t \epsilon_t, \\
             h_t^2=a_0+b_0 u_{t-1}^2  
\end{eqnarray}
We set a constant parameter for ARCH to $a_0=1$.  
Persistence parameter $b_0$ for ARCH is set to $b_0=0.3$, $0.6$, and $0.9$ in order to examine the effect of persistence of variance on the performance of the tests.  
$b_0<1$ is the necessary and sufficient condition for a weak stationarity of a semi-strong process. 
For (22), (23), and (24), the rejection frequencies of $M_a$ and $M_{wb}$ show the empirical size, because the process does not have time-varying mean. 
The performances of $V_a$, $V_b$, and $V_{wb}$ show frequencies of finding ARCH properties. 
$M_a$ overrejects the null hypothesis at 5\% significance and its over-rejection increases when the persistence parameter and/or sample size increases. 
$M_a$ is sensitive to persistence in the presence of the ARCH effect. 
$M_{wb}$ has slight over-rejection only for $b_0=0.9$ with large samples. 
However, $M_{wb}$ clearly outperforms $M_a$ and has reasonable and acceptable rejection frequencies in most cases.  
This is due to the property that the wild bootstrap can deal with heteroskedasticity of unknown form when the conditional mean is correctly specified. 
We observe the difference of power properties among $V_a$, $V_b$, and $V_{wb}$. 
$V_a$ and $V_b$ have sufficient power to find the ARCH effect. 
Their abilities to detect the ARCH effect increases for large $b_0$ and $T$. 
The ability of $V_{wb}$ is lower than that of $V_a$ and $V_b$. 
For example, the powers of $V_a$, $V_b$, and $V_{wb}$ for $b_1=0.3$ and $T=400$ are 0.937, 0.937, and 0.260, respectively.  
While the size properties of $V_{wb}$ are superior to $V_{a}$ and $V_{b}$ under a misspecified conditional mean reported in Table 2, 
the powers of $V_{wb}$ are clearly inferior to $V_{a}$ and $V_{b}$.  
The higher powers of $V_{a}$ and $V_{b}$ are due to size distortions presented in Table 2. 
The comparison indicates that the use of these tests is not effective in the viewpoint of size and power 
because the size corrected tests are needed.

We finally investigate the statistical properties of the tests for the AR process with time-varying ARCH error. 
The DGP is given by 
\begin{eqnarray}
             y_t=1 + 0.3 y_{t-1} + u_{t},\\
             u_t = h_t \epsilon_t, \\
             h_t^2=1+0.3 u_{t-1}^2 + (a_1 + b_1 u_{t-1}^2) F (\cdot), \\
             F(t, \ \gamma, \ c)= (1+ \exp \{ - \gamma (t-\frac{T}{2}) \})^{-1}-\frac{1}{2}. 
\end{eqnarray}
Time-varying parameters $(a_1,b_1)$ are set to $(a_1,b_1)=(0,0.3),$ $(0,0.6),$ $(0.5,0.3),$ and $(1,0.3)$, respectively.  
We consider DGP with $\gamma=0.01$ and $0.1.$ 
These settings are similar to those in Table 2, in which the DGP has time-varying mean. 
The results are presented in Table 4. 
$M_{wb}$ does not have over-rejection for all cases and is close to the nominal size at 5\%.  
This means that $M_{wb}$ is a reliable test for time-varying mean and does not lead to spurious time-varying mean.
Although $M_a$ has size distortions under time-varying ARCH error, 
the rejection frequencies do not depend on the size of parameters $a_1$, $b_1$, and $\gamma$ too much.  
The rejection frequencies mainly depend on sample size and slowly increase when the sample size increases. 
The results are similar to those of Zhou (2013) and Boldea, Cornea-Madeira, and Hall (2019).  
They show that the wild bootstrap is asymptotically valid for a change in mean or regression coefficients 
in the presence of conditional and unconditional heteroskedasticity$^{8}$. 
The wild bootstrap can replicate unknown heteroskedasticity of the errors. 
The property brings valid tests for linearity of the mean. 

The powers of $V_b$ are a little better than those of $V_a$.  
These tests have sufficient power to find time-varying ARCH effects. 
Furthermore, the increase in the smoothness parameter $\gamma$ brings higher power of time-varying ARCH tests. 
In particular, the effect is clear for $(a_1,b_1)=(1,0.3)$. 
When the sample size is $T=100$, 
the powers of $V_a$ and $V_b$ are 0.387 and 0.429 for $\gamma=0.01$, respectively, and 0.774 and 0.813 for $\gamma=0.1$, respectively.  
Note that the results of $V_{wb}$ are clearly different from those of $V_a$ and $V_b$. 
The powers of $V_{wb}$ are lower than those of $V_a$ and $V_b$ in all cases. 
For example, when the process has $(a_1,b_1)=(0.5,0.3)$ and $T=200$, 
the powers of $V_a$, $V_b$, and $V_{wb}$ are 0.724, 0.744, and 0.154, respectively. 
$V_a$ and $V_b$ cause large over-rejections under the null hypothesis with time-varying mean, 
whereas they have higher power under the alternative hypothesis.  
By contrast, $V_{wb}$ has better size properties than $V_a$ and $V_b$ do, 
whereas $V_{wb}$ has clearly lower power than $V_a$ and $V_b$ do. 
The Monte Carlo simulation results provide evidence that $M_{wb}$ tests yeild reliable results than do $M_a$ under ARCH and time-varying ARCH, 
and time-varying ARCH tests do not perform well in the presence of time-varying mean.

\section{Summary and conclusion}
This study examined the statistical performance of time-varying tests under misspecified conditional mean and variance.  
Although time-varying properties are frequently observed in various economic and financial data, 
it is difficult to know which time-varying property for mean or variance is present a priori. 
Researchers may employ misspecified conditional models and obtain unreliable results. 
Therefore, it is important to clarify the statistical properties of time-varying tests in misspecified conditional mean or variance models. 
Monte Carlo simulation results reveal that asymptotic tests for time-varying mean have size distortions when the variance model is misspecified, 
whereas the wild bootstrap method improves the size distortions. 
Wild bootstrap tests for time-varying mean are robust regardless of the misspecified variance, 
and can lead to reliable results of tests for time-varying mean. 
However, bootstrap tests in addition to asymptotic tests for time-varying variance have size distortions in the presence of the misspecified conditional mean.  
Tests for time-varying variance do not perform well and not provide reliable results in the presence of misspecified conditional mean. 
Robust time-varying variance tests in the presence of the misspecified conditional mean remain for further study.

\newpage
\fontsize{10pt}{20pt}\selectfont
\begin{flushleft}
{\Large Footnotes}\\
\end{flushleft} 
1. Although (1) can take a general model with $p$ lags, we consider only a single-lag model to simplify the investigation in this study. 
The same is true of variance.  \\
\\
2. Davidson and Flachaire (2008) discussed the method of the wild bootstrap, including other distributions for $\epsilon_{t}$.\\ 
\\ 
3. (12) includes conditional and unconditional heteroskedasticity. 
While conditional heteroskedasticity has $b_0>0$ and/or $b_1>0$ with $F(\cdot) \neq 0$,  
unconditional heteroskedasticity has $a_0>0$ and/or $a_1>0$ with $F(\cdot) \neq 0$ in addition to $b_0=b_1=0$. \\ 
\\ 
4. It is possible to test for the null hypothesis of $\rho_2=\rho_3=0$. 
This means that in the null hypothesis, there is an ARCH error and in the alternative hypothesis, there is a time-varying ARCH error. 
However, it is difficult to know that the variance process has ARCH properties a priori. 
When we have the null hypothesis $\rho_2=\rho_3=0$, 
we have to test for the hypothesis whether variance has ARCH error before testing for $\rho_2=\rho_3=0$. 
This takes a two-step approach and cannot test for time-varying ARCH directly. 
This may make comparisons among the tests difficult and lead to misleading results. 
Therefore, we adopt hypothesese in (14) to avoid these problems and simply test for time-varying ARCH. \\ 
\\ 
5. The test statistic is also expressed by $T R^2$, where $R^2$ is the coefficient of the determinantion of (13). $T R^2$ asymptotically has $\chi^2(3)$ distribution. 
See Gel and Chen (2012)\\
\\
6. We conduct Monte Carlo experiments under various other situations. 
For example, while tests have larger rejection frequencies when $c$ is smaller than $T/2$, 
they have smaller rejection frequencies when $c$ is larger than $T/2$.  
However, differences among the tests are similar to those of $c=T/2$. 
The results for other parameter sets $(\alpha_1,\beta_1, \gamma, \mbox{and} \ c)$ are available from the author on request.\\  
\\ 
7. See Luukkonen et al. (1988b) for similar results. \\ 
\\ 
8. Zhang and Wu (2019) proposes a nonparametric test for a change in regression coefficients in the presence of time-varying variance.

\newpage
\fontsize{11pt}{17pt}\selectfont

\newpage
\fontsize{11pt}{17pt}\selectfont
\begin{center}
Table 1: Rejection frequencies under AR with homoskedastic error 
\end{center}

\begin{center}
\begin{tabular}{c|ccccc} \hline          
                    &$M_a$   &$M_{wb}$& $V_a$ &  $V_b$  &$V_{wb}$ \\ \hline 
$\beta_0=0.3$ &           &           &           &           &             \\
$T=100$        & 0.049   & 0.051  & 0.034   &  0.046  &  0.051    \\
$T=200$        & 0.048   & 0.049  & 0.041   &  0.045   & 0.054   \\
$T=400$        & 0.053   & 0.054  & 0.040   &  0.050   & 0.051   \\  
$T=1000$       & 0.046  & 0.054  & 0.046   &  0.050  &  0.046    \\
$\beta_0=0.9$ &           &          &            &           &             \\
$T=100$        & 0.106   & 0.104  & 0.035   & 0.047   &  0.048    \\
$T=200$        & 0.080   & 0.074  & 0.043   & 0.049   &  0.044    \\
$T=400$        & 0.054   & 0.061  & 0.045   & 0.049   &  0.051    \\     
$T=1000$       &0.056   & 0.057  & 0.045   & 0.050   &  0.045     \\     
                    &           &          &            &           &              \\
    \hline
   \end{tabular}
\end{center}

\newpage
\begin{landscape}
\begin{center}
Table 2: Rejection frequencies under time-varying AR with homoskedastic error 
\end{center}

\begin{center}
\begin{tabular}{c|ccccc|ccccc} \hline
                    &\multicolumn{5}{|c|}{$\gamma=0.01$} & \multicolumn{5}{|c}{$\gamma=0.1$}   \\ \hline       
                    &$M_a$   &$M_{wb}$& $V_a$ &  $V_b$  &$V_{wb}$  & $M_a$   &$M_{wb}$& $V_a$ &  $V_b$  &$V_{wb}$\\ \hline 
$\alpha_1=0, \beta_1=0.3$ &        &           &           &     &       &           &           &           &           &            \\
$T=100$        & 0.056   & 0.061  & 0.036   &  0.043  &  0.048   & 0.380   & 0.382  & 0.036   &  0.046  &  0.042   \\
$T=200$        & 0.134   & 0.132  & 0.040   &  0.048   & 0.049  & 0.789   & 0.794  & 0.036   &  0.051   & 0.035 \\
$T=400$        & 0.625   & 0.630  & 0.049   &  0.053   & 0.041  & 0.986   & 0.988  & 0.042   &  0.067   & 0.035 \\  
$T=1000$       &   1      & 0.999  & 0.068   &  0.072  &  0.031   &   1      &    1     & 0.119   &  0.086  &  0.036  \\
$\alpha_1=0, \beta_1=0.6$ &       &           &            &     &      &         &           &            &            &           \\
$T=100$        & 0.091   & 0.090  & 0.035   & 0.043   &  0.048     & 0.943   & 0.941  & 0.055   & 0.071   &  0.029 \\
$T=200$        & 0.428   & 0.424  & 0.041   & 0.053   &  0.045     &   1       & 0.999  & 0.126   & 0.133   &  0.040  \\
$T=400$        & 0.996   & 0.997  & 0.064   & 0.076   &  0.027    &   1       &    1     & 0.286   & 0.296   &  0.083    \\     
$T=1000$       &   1      &    1     & 0.325   & 0.332   &  0.057   &  1       &    1     & 0.704   & 0.703   &  0.253      \\     
$\alpha_1=0.5, \beta_1=0.3$ &     &           &           &      &       &         &            &           &          &             \\
$T=100$        & 0.085   & 0.091  & 0.036   &  0.042  &  0.044    & 0.880   & 0.875  & 0.038   &  0.046  &  0.034  \\
$T=200$        & 0.357   & 0.365  & 0.040   &  0.047   & 0.048    & 0.998   & 0.999  & 0.049   &  0.054   & 0.045 \\
$T=400$        & 0.990   & 0.992  & 0.046   &  0.055   & 0.038   &    1      &   1     & 0.068   &  0.077   & 0.044   \\  
$T=1000$       &   1      &   1     & 0.074    &  0.078  &  0.035    &   1      &   1     & 0.139    &  0.144  &  0.070   \\
$\alpha_1=1, \beta_1=0.3$ &      &            &            &      &     &          &            &            &          &            \\
$T=100$        & 0.121   & 0.129  & 0.036   & 0.042   &  0.042    & 0.994   & 0.994  & 0.040   & 0.052   &  0.031  \\
$T=200$        & 0.694   & 0.703  & 0.036   & 0.043   &  0.044   &    1      &   1     & 0.070   & 0.080   &  0.044  \\
$T=400$        &    1      &   1     & 0.042   & 0.052   &  0.034    &    1      &   1     & 0.127   & 0.132   &  0.058  \\     
$T=1000$       &   1      &   1     & 0.119    & 0.125   &  0.040    &   1      &   1     & 0.300    & 0.311   &  0.113   \\     
                    &           &          &            &           &              &           &          &            &           &             \\
    \hline
   \end{tabular}
\end{center}
\end{landscape}

\newpage
\begin{center}
Table 3: Rejection frequencies under AR with ARCH error 
\end{center}

\begin{center}
\begin{tabular}{c|ccccc} \hline          
                    &$M_a$   &$M_{wb}$& $V_a$ &  $V_b$  &$V_{wb}$ \\ \hline 
$b_0=0.3$ &           &           &           &           &             \\
$T=100$        & 0.079   & 0.049  & 0.372   &  0.407  &  0.065    \\
$T=200$        & 0.086   & 0.052  & 0.681   &  0.705   & 0.120   \\
$T=400$        & 0.098   & 0.052  & 0.937   &  0.937   & 0.260   \\  
$T=1000$       & 0.100  & 0.053  & 0.999   &  0.999  &  0.558    \\
$b_0=0.6$ &           &          &            &           &             \\
$T=100$        & 0.107   & 0.049  & 0.686   & 0.719   &  0.184    \\
$T=200$        & 0.144   & 0.057  & 0.946   & 0.956   &  0.323    \\
$T=400$        & 0.191   & 0.051  & 0.998   & 0.999   &  0.498    \\     
$T=1000$       &0.231   & 0.055  &    1      &    1      &  0.674     \\    
$b_0=0.9$ &           &          &            &           &             \\
$T=100$        & 0.129   & 0.055  & 0.802   & 0.841   &  0.253    \\
$T=200$        & 0.201   & 0.057  & 0.976   & 0.985   &  0.366    \\
$T=400$        & 0.282   & 0.062  & 0.998   & 0.999   &  0.474    \\     
$T=1000$       &0.401   & 0.060  & 0.999   &     1     &  0.561     \\    
                    &           &          &            &           &              \\
    \hline
   \end{tabular}
\end{center}

\newpage
\begin{landscape}
\begin{center}
Table 4: Rejection frequencies under AR with time-varying ARCH error 
\end{center}

\begin{center}
\begin{tabular}{c|ccccc|ccccc} \hline
                    &\multicolumn{5}{|c|}{$\gamma=0.01$} & \multicolumn{5}{|c}{$\gamma=0.1$}   \\ \hline       
                    &$M_a$   &$M_{wb}$& $V_a$ &  $V_b$  &$V_{wb}$  & $M_a$   &$M_{wb}$& $V_a$ &  $V_b$  &$V_{wb}$\\ \hline 
$a_1=0, b_1=0.3$ &        &           &           &     &       &           &           &           &           &                         \\
$T=100$        & 0.079   & 0.049  & 0.381   &  0.407  &  0.071    & 0.072   & 0.049  & 0.418   &  0.454  &  0.098   \\
$T=200$        & 0.091   & 0.055  & 0.692   &  0.706   & 0.128   & 0.088    & 0.053  & 0.738   &  0.756   & 0.175   \\
$T=400$        & 0.101   & 0.051  & 0.946   &  0.950   & 0.276   & 0.105    & 0.054  & 0.962   &  0.963   & 0.329   \\  
$T=1000$       & 0.116   & 0.056  & 0.999   &  0.999  &  0.576   & 0.117   & 0.050   & 0.999   &    1     &  0.597   \\
$a_1=0, b_1=0.6$ &       &           &            &     &      &         &           &            &            &                          \\
$T=100$        & 0.082   & 0.051  & 0.384   & 0.414   &  0.069     & 0.079   & 0.059  & 0.521   & 0.560   &  0.153  \\
$T=200$        & 0.091   & 0.055  & 0.712   & 0.733   &  0.144     & 0.095   & 0.059  & 0.863   & 0.875   &  0.298  \\
$T=400$        & 0.109   & 0.053  & 0.962   & 0.970   &  0.333    &  0.115   & 0.056  & 0.991   & 0.991   &  0.458  \\     
$T=1000$       & 0.151  &  0.054  &   1      &   1      &  0.627    &  0.146    & 0.061 &    1      &    1      &  0.645  \\     
$a_1=0.5, b_1=0.3$ &     &           &           &      &       &         &            &           &          &                          \\
$T=100$        & 0.077   & 0.054  & 0.381   &  0.414  &  0.071    & 0.072    & 0.055  & 0.563  &  0.600  &  0.202   \\
$T=200$        & 0.086   & 0.054  & 0.724   &  0.744   & 0.154     & 0.083   & 0.052  & 0.910   &  0.922   & 0.384  \\
$T=400$        & 0.102   & 0.056  & 0.976   &  0.981   & 0.385    &  0.096   & 0.050  & 0.998   &  0.998   & 0.570  \\  
$T=1000$       & 0.125   & 0.055  &   1      &    1      &  0.725    &  0.118   & 0.052  &   1      &    1      &  0.746  \\
$a_1=1, b_1=0.3$ &       &            &            &      &     &          &            &            &          &                         \\
$T=100$        & 0.073   & 0.054  & 0.387   & 0.429   &  0.082    & 0.074   & 0.053   & 0.774   & 0.813   &  0.410  \\
$T=200$        & 0.091   & 0.055  & 0.752   & 0.790   &  0.196   &  0.081    & 0.057  & 0.992   & 0.995   &  0.595  \\
$T=400$        & 0.101   & 0.054  & 0.989   & 0.996   &  0.539    &  0.090   & 0.055  &    1      &    1      &  0.723  \\     
$T=1000$       &0.118   & 0.052  &   1       &    1      &  0.813    &  0.117   & 0.053  &    1      &    1     &  0.816   \\     
                    &           &          &            &           &              &           &          &            &            &            \\
    \hline
   \end{tabular}
\end{center}
\end{landscape}

\newpage
\fontsize{10pt}{10pt}\selectfont
\begin{figure}
\begin{center}
\caption{Value of the transition function with $T=200$}
   \includegraphics[width=13cm,height=7cm,clip]{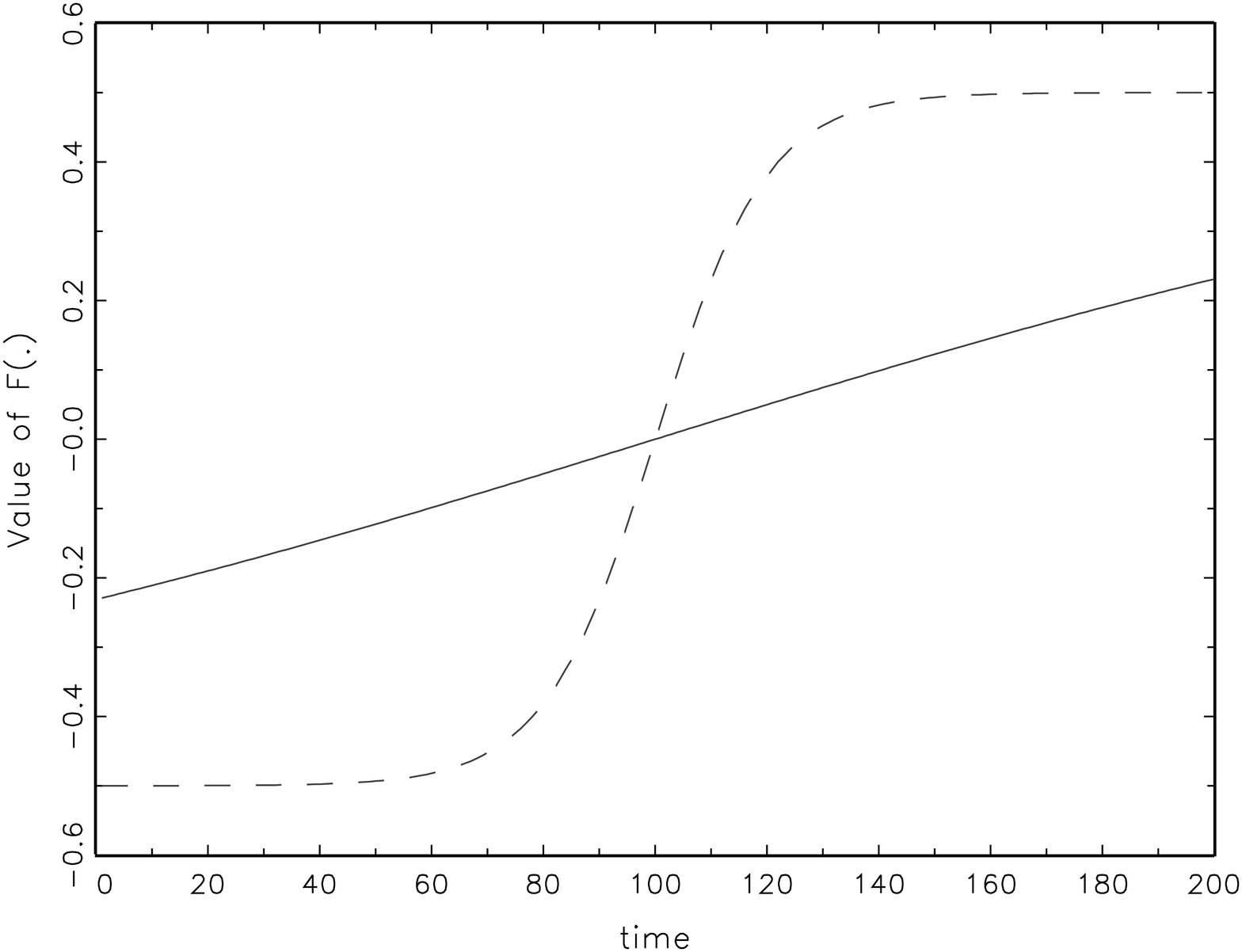} \\
\caption{Value of the transition function with $T=1,000$}
   \includegraphics[width=13cm,height=7cm,clip]{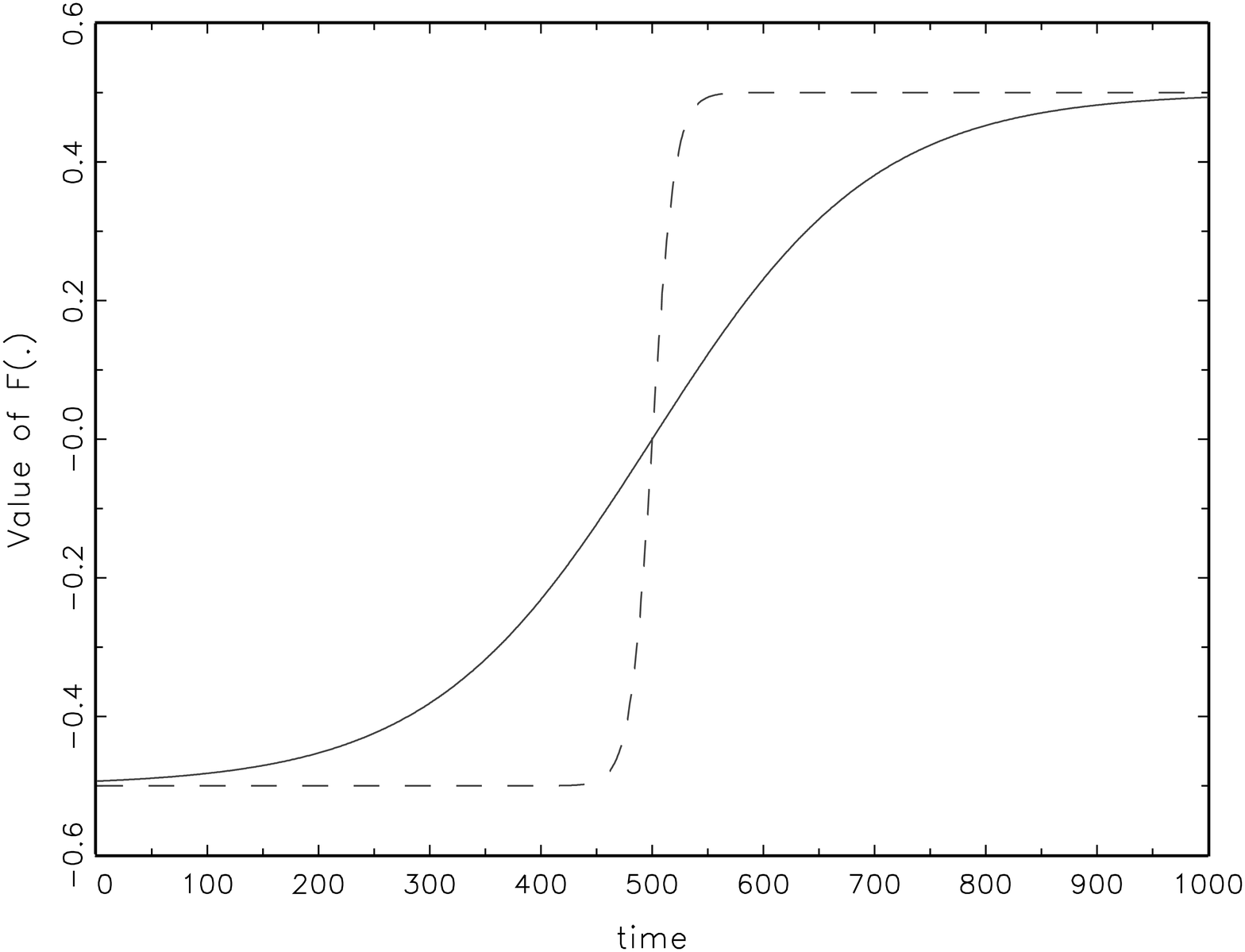} \\
Solid line has $\gamma=0.01$. Dashed line has $\gamma=0.1$.
\end{center}  
\end{figure}

\end{document}